\documentclass[journal=aamick,manuscript=letter,layout=twocolumn]{achemso}

\usepackage[version=3]{mhchem} 
\usepackage{graphicx}
\usepackage{dcolumn}
\usepackage{bm}
\usepackage[mathlines]{lineno}
\usepackage{amssymb}
\usepackage{amsmath}
\usepackage{soul,color}
\usepackage[colorlinks,linkcolor=blue,anchorcolor=blue,citecolor=blue,urlcolor=blue]{hyperref}
\usepackage{dcolumn}

\usepackage{threeparttable}  
\usepackage{cmll}
\usepackage{multirow}
\usepackage{booktabs}
\usepackage{color}
\usepackage{datetime}

\title{Thermoelectric Penta-Silicene with a High Room-Temperature Figure of Merit}

\author{Zhibin Gao}
\email{zhibin.gao@nus.edu.sg}
\affiliation{Department of Physics, National University of Singapore,
             Singapore 117551, Republic of Singapore}
\author{Jian-Sheng Wang}
\affiliation{Department of Physics, National University of Singapore,
             Singapore 117551, Republic of Singapore}

\date{\today}


\begin{document}


\begin{abstract}
Silicon is one of the most frequently used chemical elements of the periodic
table in nanotechnology\cite{goodilin2019nanotechnology}. Two-dimensional (2D)
silicene, a silicon analog of graphene, has been readily obtained to make
field-effect transistors since 2015\cite{tao2015silicene,tsai2013gated}.
Recently, as new members of the silicene family, penta-silicene and its
nanoribbon have been experimentally grown on Ag(110) surface with exotic
electronic properties\cite{cerda2016unveiling,prevot2016si,sheng2018pentagonal}.
However, the thermoelectric performance of penta-silicene has not been so far
studied that would hinder its potential applications of electric generation
from waste heat and solid-state Peltier coolers. Based on Boltzmann transport
theory and \textit{ab initio} calculations, we find that penta-silicene shows
a remarkable room temperature figure of merit \textit{ZT} of 3.4 and 3.0 at the
reachable hole and electron concentrations, respectively. We attribute this high
\textit{ZT} to the superior ``pudding-mold'' electronic band structure and ultralow
lattice thermal conductivity. The discovery provides new insight into the transport
property of pentagonal nanostructures and highlights the potential applications of
thermoelectric materials at room temperature.
\end{abstract}

\section*{Introduction}
The thermoelectric effect directly converts waste heat to electrical energy.
Its efficiency is defined by a dimensionless figure of merit \textit{ZT},
written as
$ ZT = S^2 \sigma T / (\kappa_e + \kappa_L) $ in which \textit{S}, $\sigma$,
$\kappa_e$, $\kappa_L$ are the Seebeck coefficient, electrical conductivity,
electronic thermal conductivity and lattice thermal conductivity, respectively.
Since \textit{S}, $\sigma$, and $\kappa_e$ are closely entangled with each other,
it is challenging to increase numerator and decrease denominator concurrently,
gaining a high \textit{ZT}. Explicitly, an average \textit{ZT} larger than 2 would
make thermoelectric materials without toxic substances attractive and competitive
when compared with other types of energy conversion\cite{he2017advances}.
Furthermore, taking into account the application in daily life, near room
temperature thermoelectric materials with high \textit{ZT} are desirably needed.

A reduced $\kappa_L$ can improve the thermoelectric efficiency,
\textit{ZT}\cite{snyder2008complex}.
%
One can use resonant bonding\cite{lee2014resonant}, and lone electron
pairs\cite{nielsen2013lone} to reduce $\kappa_L$.
Besides, the power factor \textit{S}$^2$$\sigma$ can be augmented by electronic
band engineering\cite{heremans2008enhancement,pei2011convergence,liu2012convergence}
and quantum confinement effect\cite{dresselhaus2007new}.
Furthermore, 2D materials have inherent advantages to moderate the contradiction
between \textit{S} and $\sigma$ due to the tunable electronic band gap, resulting
in a high power factor. This is much simpler than in the case of bulk materials.
In the case of 2D materials, one has great potential to break the
complicated relationship between \textit{S} and $\sigma$ due to the quantum
confinement effect. This argument was firstly given by pioneer Mildred
Dresselhaus\cite{dresselhaus2007new}. She also pointed out that one of the
advantages of 2D layered materials is that one can actually engineer the band gap.
In 2D, one can simply change the number of layers or apply stress/strain to modify
the band gap, or even tune the chemical bond to modify the electronic band
gap\cite{dresselhaus2007new}.

In 2014, a theoretical work predicted a new 2D carbon allotrope called penta-graphene
with favorable stability\cite{zhang2015penta} whose prototype was firstly proposed
in iron-based material with exotic magnetic frustration\cite{ressouche2009magnetic}.
Afterward, the pentagonal system has attracted much attention, such as unexpected thermal
conductivity\cite{wu2016hydrogenation,liu2016disparate,gao2019ultralow},
and seamless electrical
contacts\cite{oyedele2019defect}. Recently, a stable penta-silicene was found by reducing
the Coulomb interaction of silicon dimers\cite{guo2019lattice}. They also reported that
it has ultrahigh Curie temperature of 1190~K. However, the thermoelectric performance of
penta-silicene is still lacking.

In this study, we explore the thermoelectric property of penta-silicene based on the
Boltzmann transport theory and \textit{ab initio} calculations. It has good thermal,
dynamical, and mechanical stability compared with many typical 2D materials. We find
that penta-silicene has a nearly direct band gap of 0.68 eV at the DFT-HSE06 level.
Lattice thermal conductivities of penta-silicene have values of 1.66~W/mK and 1.29~W/mK
along x- and y-axis, respectively. At room temperature, Penta-silicene shows a maximum
figure of merit \textit{ZT} of 3.4 and 3.0 at the reachable hole and electron
concentrations. Our work indicates that penta-silicene is a promising thermoelectric
material, especially near room temperature.

\section*{Computational methods}
We perform DFT calculations using projector-augmented-wave (PAW)
method\cite{blochl1994projector,kresse1999ultrasoft},
Perdew-Burke-Ernzerhof (PBE)\cite{perdew1996generalized} and
hybrid exchange-correlation HSE06
functional\cite{heyd2003hybrid,krukau2006influence} with default
mixing parameter value $\alpha = 0.25 $ in the VASP
code\cite{kresse1996efficient,kresse1996efficiency,kresse1994ab}.
Plane waves with 550~eV kinetic cutoff energy are used. The vacuum
distance between the neighboring layer is set to be 20 \AA ~removing
the nonphysical long-range electrostatic interaction. The ionic
Hellmann-Feynman forces in each atom and totally free energy are
converged to  10$^{-4}$ eV/ \AA ~and 10$^{-8}$ eV in the structure
optimization and band calculation. The Brillouin zone is sampled by
uniform 21 $\times$ 21 $\times$ 1. The electronic transport properties
are calculated using the electronic Boltzmann transport theory
implemented in BoltzTraP\cite{madsen2006}.
In the phonon calculation, we used 5 $\times$ 5 $\times$ 1
supercell and 2 $\times$ 2 $\times$ 1 k-point sampling to compute the
second and third-order force constants. To solve the phonon Boltzmann
transport equation, we adopted a 101 $\times$ 101 $\times$ 1
\textit{$\Gamma$}-centered q-grid. We also have tested the convergence
of lattice thermal conductivity with respect to the cutoff radius,
shown in the Supporting Information. The linearized phonon Boltzmann
transport equation is solved by ShengBTE\cite{ShengBTE2014} via a full
iteration.

\section*{Results}
\subsection*{Crystal structure}

\begin{table*}
 \begin{threeparttable}
		\caption{The calculated physical properties of penta-graphene
                 and penta-silicene.
                 $\left|\vec{a_1}\right|$($\left|\vec{a_2}\right|$)
                 is the lattice constant. \textit{d} and \textit{h}
                 are the intrinsic buckling distance and effective
                 thickness. \textit{E$_c$} is the cohesive energy
                 per atom. \textit{C$_{ij}$} is the elastic modulus
                 tensor. \textit{G} and $\nu$ are the shear modulus
                 and Poisson's ratio.}\label{Tab1}
    		\renewcommand\arraystretch{1.5}
		\begin{tabular*}{1.0\textwidth}{p{2.1cm}p{1.5cm}p{1.5cm}p{1.3cm}p{1.85cm}p{1.5cm}p{1.5cm}p{1.3cm}p{1.5cm}}		
			\hline \hline
	        Materials & $\left|\vec{a_1}\right|$($\left|\vec{a_2}\right|$) & \textit{d} & \textit{h} & \textit{E$_c$} & \textit{C$_{11}$} & \textit{C$_{12}$} & \textit{G} & $\nu$ \\
            Unit         & ~~{\AA} & {\AA} & {\AA} & (eV/atom) & (GPa) & (GPa) & (GPa) & --- \\
			\hline
			penta-Gr     & 3.64  & 1.21  &  4.61  & 7.08   & 584.34  & -46.88  & 328.94  & -0.09	 \\
			penta-Si     & 5.58  & 2.44  &  6.64  & 3.92   & 43.13   & -23.56  & 30.31   & -0.55 	 \\
			\hline \hline
		\end{tabular*}
    \end{threeparttable}
\end{table*}

\begin{figure*}[t!]
\includegraphics[width=2.0\columnwidth]{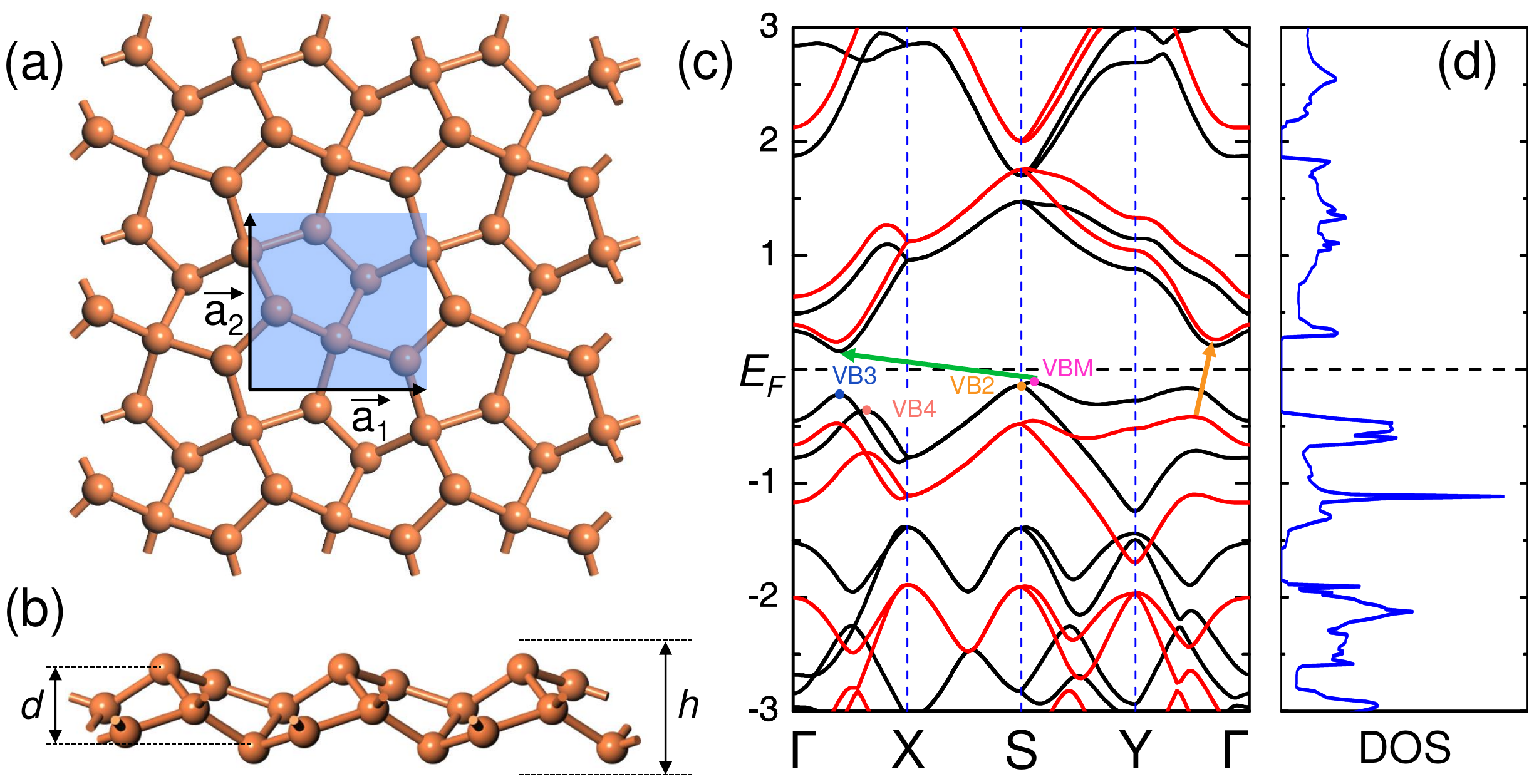}
\caption{(a,b) Top and side views of the atomic structure of the
monolayer penta-silicene in a 3$\times$3 supercell.
(c) Electronic band structure of penta-silicene using the
DFT-PBE (black) and DFT-HSE06 (red) functionals. (d) Electronic
density of states (DOS) at the HSE06 level. The indirect fundamental
band gaps are marked by the green (PBE) and orange (HSE06) arrows in
(c). The Fermi levels are set to zero. The sum of the intrinsic
thickness \textit{d} and two van der Waals radii of the outermost
atom is defined as the effective thickness \textit{h}. The
primitive cell is indicated by blue shading in (a). Four hole
pockets in valence bands around the Fermi level are labeled
by different colors and numbers (VBM, VB2, VB3, and VB4). VBM
denotes the valence band maximum. In the first
Brillouin zone, the high symmetry \textit{k} points are:
\textit{$\Gamma$}(0 0 0), \textit{X}(1/2 0 0),
\textit{S}(1/2 1/2 0) and \textit{Y}(0 1/2 0), respectively.
}\label{fig1}
\end{figure*}

The optimized monolayer penta-silicene is shown in Figure~\ref{fig1}a,b.
There are 6 atoms in the primitive cell indicated
by blue shading. From the top view, penta-silicene is the same as
penta-graphene\cite{zhang2015penta}. But there is a little distortion
of penta-silicene compared with penta-graphene\cite{guo2019lattice}.
This is due to the tilting of silicon dimers in the penta-silicene
in order to reduce the strong Coulomb repulsion and stabilize the
crystal structure. It means that one cannot obtain the penta-silicene
by simple element substitution. One should break the symmetry by
moving one of the atoms in order to further lower the energy
and reach a local minimum potential profile. The symmetry becomes
\textit{P}2$_1$ (space group no.~4) from
\textit{P}42$_1$\textit{m} (space group no.~113). The optimized
lattice constants of penta-silicene and penta-graphene, shown in
Table ~\ref{Tab1}, are
$\left|\vec{a_1}\right|$ = $\left|\vec{a_2}\right|$ = 5.58 {\AA}
and 3.64 {\AA}, respectively, which are in good agreement with
previous results\cite{guo2019lattice,zhang2015penta}. It is
reasonable due to the augment of atomic size in the same main
group IV. Intrinsically buckling distance also increases from
1.21 {\AA} to 2.44 {\AA} from penta-graphene to penta-silicene.
The cohesive energy can be calculated by $ E_c = (n*E_{Si}-E_0)/n $
in which \textit{E}$_{Si}$, \textit{E}$_0$, and \textit{n} are the
energy of single silicon atom, the energy of the whole system at
equilibrium state, and the number of atoms in the system. The calculated
\textit{E}$_c$ of penta-graphene and penta-silicene are 7.08 eV/atom
and 3.92 eV/atom. Note that, in the experiment, \textit{E}$_c$ of the
experimentally accessible silicene (hexagonal symmetry with 2 atoms in
the primitive cell) and phosphorene are
3.71~eV/atom\cite{fleurence2012experimental,feng2012evidence} and
3.61~eV/atom\cite{liu2014phosphorene,li2014black}, respectively.
This indicates that the stability of penta-silicene is comparable with
hexagonal silicene and black phosphorene, suggesting a robust chemical
environment bond to maintain the stability of penta-silicene.


In 3D materials, the elastic tensor is a 6 $\times$ 6 tensor
with 36 number of tensor elements. By considering the symmetry of the
crystal structure, 6 $\times$ 6 tensor can be further simplified according
to the specific crystal systems. There are 7 crystal systems in solid.
The higher the symmetry of the crystal system, the smaller the number of
independent tensor elements. For 2D materials, The elastic tensor is a
3 $\times$ 3 tensor. To be specific, the mechanical property has a close
relationship with phonon modes on the crystal momentum near the center of
the Brillouin zone\cite{liu2016continuum}.
For penta-silicene,
there are 3 independent elastic components which are \textit{C}$_{11}$,
\textit{C}$_{12}$ and \textit{C}$_{66}$. Note that the value of
\textit{C}$_{66}$ is equal to the shear modulus \textit{G}. The calculated
mechanical data of penta-silicene is shown in Table ~\ref{Tab1}. The elastic
constants of any stable 2D materials must satisfy
\textit{C}$_{11}$\textit{C}$_{22}$ $-$ \textit{C}$^2_{12}$ $>$ 0 and
\textit{C}$_{66}$ $ > $ 0. From the table, it can be found that penta-silicene
is mechanically stable. Furthermore, the value of \textit{C}$_{12}$ is
$-46.88$~GPa, indicating a negative in-plane Poisson's ratio that is defined
as\cite{gao2017novel} $\nu_{xy} = \frac{\textit{C}_{21}} {\textit{C}_{22}}$ and
$\nu_{yx} = \frac{\textit{C}_{12}} {\textit{C}_{11}}$. For penta-graphene and
penta-silicene, \textit{C}$_{11}$ $=$ \textit{C}$_{22}$ and
\textit{C}$_{12}$ $=$ \textit{C}$_{21}$. \textit{C}$_{11}$ of
penta-silicene is smaller than that of Si-\textit{Cmma} of and
Si-\textit{Pmma}\cite{zhou2019si}, indicating a more flexible mechanical
properties than other silicon allotropes. The calculated $\nu$ of penta-graphene
and penta-silicene are $-0.09$ and $-0.55$. The value of $\nu$ in penta-silicene
is $5$ times larger than that of penta-graphene. This mechanical property of
negative Poisson's ratio is highly desirable for shock absorption in
transistors\cite{gao2017novel,gao2018two}.

\subsection*{Electronic band structure}
The electronic band structure of penta-silicene is shown in
Figure~\ref{fig1}c,d. At PBE level (black), the indirect
band gap indicated by the green arrow is 0.28~eV. Since the fundamental
band gap is usually underestimated in DFT-PBE calculations, we also
calculate it using HSE06 functional (red). Overall, The entire energy
bands are almost unchanged except for the shift up of the conduction
bands and the shift down of the valence bands. Interestingly, the HSE06
band shows a nearly direct band gap of 0.68~eV indicated by the orange
arrow. This change from the indirect-to-direct band gap would
significantly enhance the optical absorbance\cite{zhu2017multivalency}.
Some top valence bands of monolayer penta-silicene around the Fermi
level not only are quite close, but also are degenerate energetically.
The conduction bands around the Fermi level have no similar behavior.
We marked the ``mountain peaks'' of valence bands by different colors
and numbers (VBM, VB2, VB3, and VB4). As a matter of fact, these
degenerate valence bands will significantly enhance DOS and further
Seebeck coefficient\cite{guo2015first,gao2018high} that directly enters
the final figure of merit \textit{ZT}. Furthermore, since the valence
bands are highly degenerate, while conduction bands are non-degenerate,
this leads to an asymmetric DOS in Figure~\ref{fig1}d. This physical
picture is known as ``pudding-mold'' and desirably required to achieve
a large \textit{ZT}\cite{heremans2008enhancement}. We will further
discuss it in the following.

\subsection*{Electronic transport properties}

\begin{figure*}[t!]
\includegraphics[width=2.0\columnwidth]{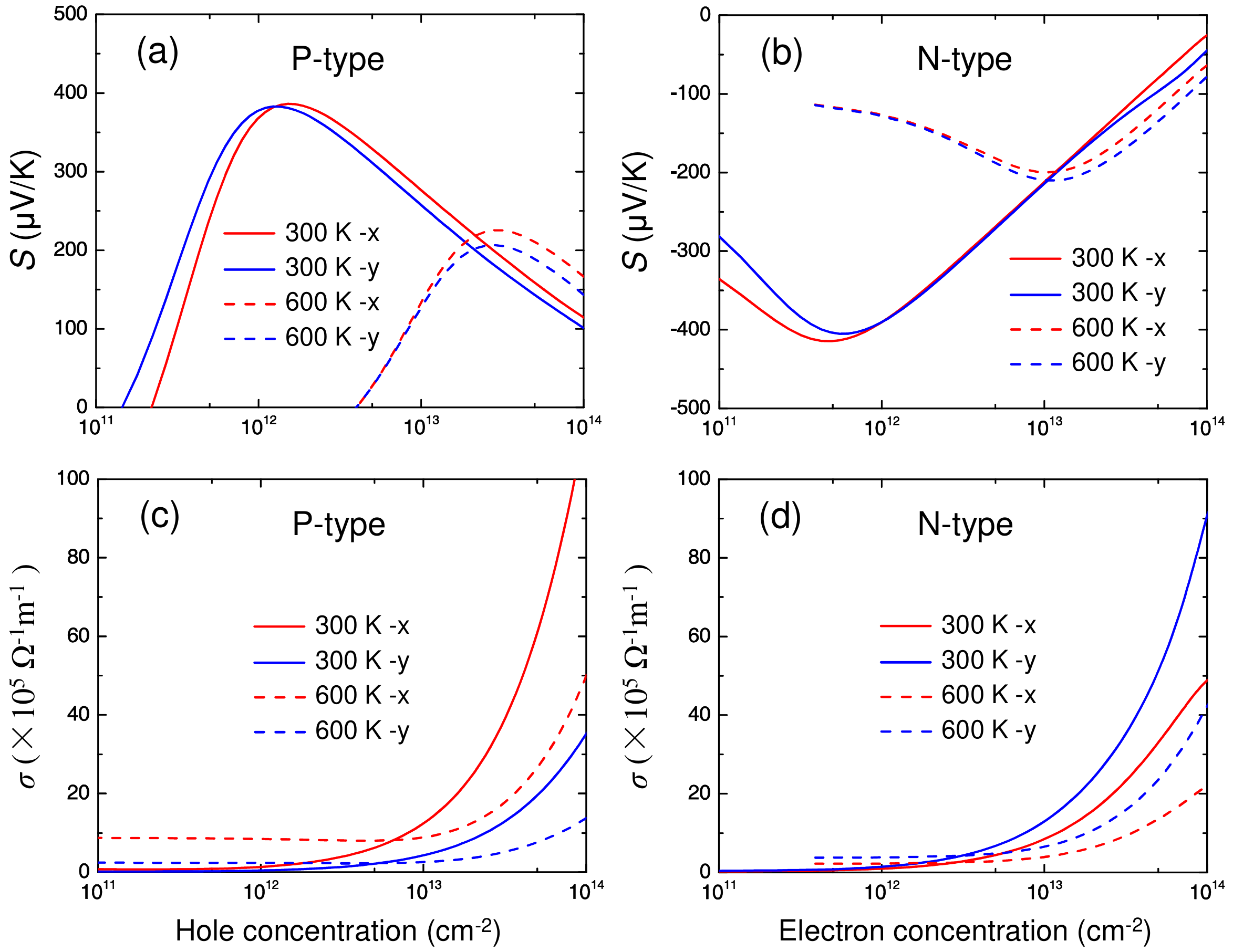}
\caption{Calculated temperature-dependent electronic transport
coefficients. (a,b) Seebeck coefficient \textit{S} and (c,d)
electrical conductivity $\sigma$ as a function of temperature
and carrier concentration (holes and electrons) for x- and
y-axis at 300 K and 600 K.
}\label{fig2}
\end{figure*}

In order to obtain the electronic transport properties, such as
Seebeck coefficient \textit{S}, electrical conductivity $\sigma$, and electronic
thermal conductivity $\kappa_e$, one has to compute the following quantities
based on the Boltzmann transport theory, written as a function of the tensor
\textbf{K$_n$}\cite{madsen2006,gao2018high}

\begin{equation} %
\begin{aligned}
\label{tensor}
\textbf{K$_n$}=&\frac{2} {(2 \pi)^3} \sum_{i} \int d^3 \textbf{k}  \tau_i(\textbf{k}) \textbf{v}_i(\textbf{k})
\otimes \textbf{v}_i(\textbf{k}) \\
&\times [\varepsilon_i(\textbf{k}) - \mu ]^n [-\frac{\partial f(\mu, T, \varepsilon_i)} {\partial \varepsilon_i}],
\end{aligned}
\end{equation}
\begin{equation} %
\label{sigma}
\sigma = e^2 \textbf{K$_0$},
\end{equation}
\begin{equation} %
\label{seebeck}
S = \frac{1} {e T} \textbf{K$_1$} \textbf{K$_0$}^{-1},
\end{equation}
\begin{equation} %
\label{electronic_kaapa}
\textit{k$_e$} =  \frac{1} {T} (\textbf{K}_2 - \textbf{K}_1^2 \textbf{K}_0^{-1}), \\
\end{equation}
in which $2$ is for the spin degeneracy, $\textbf{v}_i(\textbf{k})$ is
the electron group velocity of the wave vector \textbf{k} and band index
\textit{i}. $\varepsilon_i$, $\mu$, and $\tau_i(\textbf{k})$ are
electronic energy, chemical potential, and electronic relaxation time.
\textit{V} is the volume. Alternatively, $\kappa_e$ can also be
calculated through the Wiedemann–Franz law\cite{madsen2006,gao2018high}
\begin{equation} %
\label{electronic_kaapa2}
\textit{k$_e$} = L \sigma T,
\end{equation}
where $L$ is a constant called Lorenz number with a value of
2.4 $\times$ 10$^{-8}$ W $\Omega$ K$^{-2}$. The calculated \textit{k$_e$}
from Eq.~(\ref{electronic_kaapa}) and Eq.~(\ref{electronic_kaapa2}) are
the same for semiconductors, which has been verified by many previous
works\cite{li2019improved,toberer2012advances,wang2017exceptional}.

Here we consider two temperatures (300 K and 600 K) in the
following calculations. The calculated \textit{S} and $\sigma$ are shown
in Figure~\ref{fig2}. For both p-type and n-type doping, \textit{S}
firstly increases (absolute value) at low carrier
concentration (\textit{n} $<$ 10$^{12}$ cm$^{-2}$), and then linearly
decreases at high carrier
concentration (10$^{12}$ cm$^{-2}$ $<$ \textit{n} $<$ 10$^{14}$ cm$^{-2}$).
According to the Mahan-Sofo theory, \textit{S} for degenerate 2D
semiconductors can be written as\cite{snyder2008complex}
\begin{equation} %
\label{seebeck2}
\textit{S$_{2D}$} = \frac{2 \pi^3 k_B^2 T} {3 e h^2 n} m^*_d,
\end{equation}
where \textit{h}, \textit{n}, and m$^*_d$ are the Planck constant, carrier
concentration and DOS effective mass around \textit{E$_F$}. Since electrons
of 2D materials only have freedom in the plane, \textit{S$_{2D}$} is quite
different from
\textit{S$_{3D}$} = $\frac{ 8 \pi^2 k_B^2 T } { 3 q h^2 }$ $(\frac{\pi} {3n})^{(2/3)}$ $m^*_d$
in bulk materials\cite{gao2018high}. Hence, there is a competition relation
between \textit{n} and \textit{$m^*_d$}. For high concentration, our calculated
\textit{S$_{2D}$} of penta-silicene is inversely proportional to \textit{n},
which is in a good agreement with the Mahan-Sofo theory. For low concentration,
\textit{S$_{2D}$} of penta-silicene is induced by the bipolar effect in narrow
band gap semiconductors\cite{bahk2016minority,gao2018high}. Interestingly,
\textit{S$_{2D}$} of penta-silicene at 300 K is generally larger than that of
in 600 K. The largest \textit{S$_{2D}$} of penta-silicene has a value of
400 $\mu$ V/K at 10$^{12}$ cm$^{-2}$ carrier concentration for both p- and
n-types, which is double of 200 $\mu$ V/K of 2D SnSe\cite{chang20183d} at the
same condition.

\begin{table*}
 \begin{threeparttable}
		\caption{The calculated effective masses \textit{m$_x^*$/m$_0$}
                 and \textit{m$_y^*$/m$_0$}, deformation potential
                 constants \textit{E$_x$} and \textit{E$_y$}, 2D elastic
                 modulus C$_x^{2D}$ and C$_y^{2D}$, and carrier mobilities
                 $\mu_x^{2D}$ and $\mu_y^{2D}$ based on Eq.~(\ref{mobility})
                 for x- and y-axis at 300 K.}\label{Tab2}
    		\renewcommand\arraystretch{1.5}
		\begin{tabular*}{1.0\textwidth}{p{2.3cm}p{1.5cm}p{1.5cm}p{1.3cm}p{1.3cm}p{1.5cm}p{1.5cm}p{1.5cm}p{1.5cm}}		
			\hline \hline
	        Carrier type  & \textit{m$_x^*$/m$_0$} & \textit{m$_y^*$/m$_0$} & \textit{E$_x$} & \textit{E$_y$} & C$_x^{2D}$ & C$_y^{2D}$  & $\mu_x^{2D}$ & $\mu_y^{2D}$ \\
                          & (G-X) & (G-Y) & (eV) & (eV) & (J m$^{-2}$)  & (J m$^{-2}$) & m$^2$V$^{-1}$s$^{-1}$ & m$^2$V$^{-1}$s$^{-1}$ \\
			\hline
			electron      & 0.254   & 0.430   & 3.802   & 2.745   & 94.753	& 92.851   & 0.166   & 0.185	 \\
			hole          & 0.224   & 1.106   & 1.546   & 2.737   & 94.753	& 92.851   & 0.756   & 0.048	 \\
			\hline \hline
		\end{tabular*}
    \end{threeparttable}
\end{table*}

\begin{figure*}[t!]
\includegraphics[width=2.0\columnwidth]{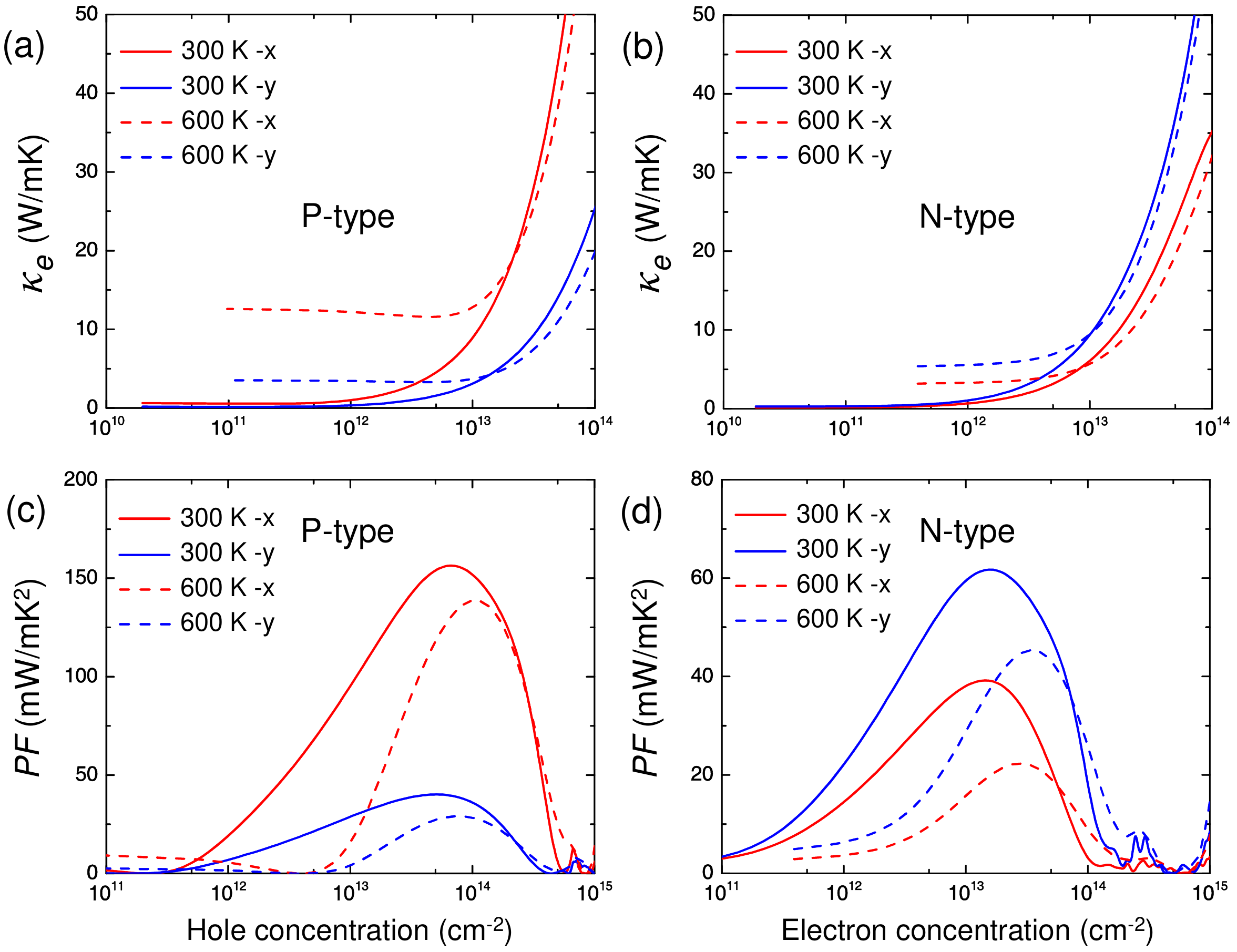}
\caption{Calculated temperature-dependent electronic transport
coefficients of penta-silicene. (a,b) electronic thermal
conductivity ($\kappa_e$) and (c,d) power factor
\textit{S}$^2\sigma$ (\textit{PF}) as a function of temperature
and carrier concentration (holes and electrons) along x- and
y-axis at 300 K and 600 K.}\label{fig3}
\end{figure*}

\begin{figure*}[t!]
\includegraphics[width=2.0\columnwidth]{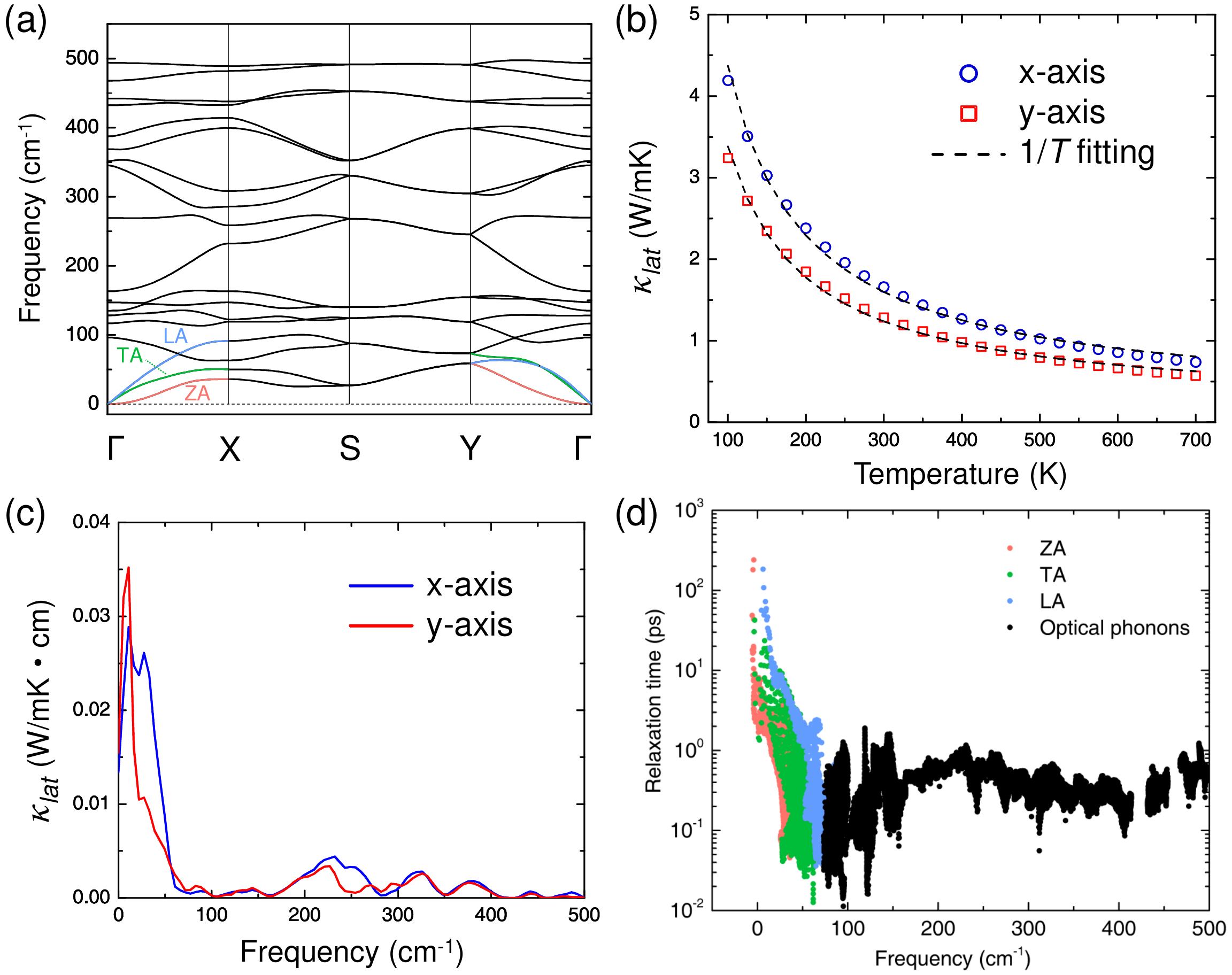}
\caption{Lattice heat transport properties of penta-silicene. (a)
phonon dispersion, (b) lattice thermal conductivity ($\kappa_L$)
as a function of temperature, (c) frequency-resolved $\kappa_L$
along x- and y-axis at 300 K, and (d) three-phonon relaxation
time at room temperature with phonon modes resolution. The
acoustic phonon branches (ZA, TA, and LA) are indicated in
different colors.}\label{fig4}
\end{figure*}

\begin{table*}
 \begin{threeparttable}
		\caption{The calculated electronic relaxation times $\tau_e$
                 and $\tau_h$, sound velocities $\upsilon_{TA}$ and
                 $\upsilon_{LA}$, Debye temperatures $\Theta_{TA}$
                 and $\Theta_{LA}$ with definition of
                 $\Theta_D = h \omega/k_B$ associated with the
                 maximum acoustic phonon frequency, and maximum figure
                 of merits \textit{ZT}$_e$ and \textit{ZT}$_h$ along
                 x- and y-axis at 300 K. The subscripts ``e'' and
                 ``h'' indicate electron and hole, respectively.}\label{Tab3}
    		\renewcommand\arraystretch{1.5}
		\begin{tabular*}{1.0\textwidth}{p{2.3cm}p{1.5cm}p{1.5cm}p{1.3cm}p{1.3cm}p{1.5cm}p{1.5cm}p{1.5cm}p{1.5cm}}		
			\hline \hline
	        Direction  & $\tau_e$ & $\tau_h$ & $\upsilon_{TA}$ & $\upsilon_{LA}$ & $\Theta_{TA}$ & $\Theta_{LA}$ & \textit{ZT}$_e$ & \textit{ZT}$_h$ \\
                          & (ps) & (ps) & (km/s) & (km/s) & (K)  & (K) & @300 K  & @300 K \\
			\hline
			G-X      & 0.240   & 0.964   & 3.99  & 4.77   & 72.32  & 131.51  & 2.18   & 3.43	 \\
			G-Y      & 0.453   & 0.302   & 4.70  & 5.54   & 91.57  & 105.78  & 3.04   & 2.24	 \\
			\hline \hline
		\end{tabular*}
    \end{threeparttable}
\end{table*}

Since the output electrical conductivity $\sigma$ from the conventional
Boltzmann transport theory is closely dependent on relaxation time $\tau$,
a method to evaluate $\tau$ should be applied appropriately. Considering
$\tau$ in many materials generally have an order of $10 \sim 12$~ps, some
works use a constant number to avoid this dilemma\cite{bilc2015,he2016}.
As a matter of fact, there are many factors to impact $\tau$, such as
acoustic phonons, nonpolar optical phonons, and ionized impurities. $\tau$
in different situations has different expressions\cite{popescu2009model}.
As an accepted rule of thumb, acoustic phonons play the most important role
in $\tau$ and the acoustic phonon limited carrier mobility $\mu$ using
deformation potential (DP) theory of 2D materials can be written as\cite{gao2018high}
\begin{equation} %
\label{mobility}
\mu_{2D} = \frac{e \hbar^3 C_{2D} } {k_B T m^* m_d^* E_i^2 }, \\
\end{equation}
%
in which the parameter can be easily found elsewhere\cite{gao2018high,qiao2014}.
The final relaxation time has a relation with mobility:
$\tau$ = $\frac{ m^* \mu } { e }$. The calculated data is shown in Table ~\ref{Tab2}.
Note that here we use single parabolic band model from the original DP
theory, which works very well for non-degenerated electronic band structure. Since
penta-silicene has a distortion compared with penta-graphene, the effective
mass \textit{m$^*$}, and deformation potential constant \textit{E} show a clear
anisotropic behavior. Note that in the calculation of \textit{E$_x$} and \textit{E$_y$},
the energy of hole (electron) must be shifted with respect to the vacuum energy with an
expression of
\textit{E}$^{h, e}$ - \textit{E}$_{vac}$ $\propto$ \textit{E}$^{h,e}$ $\cdot$ ($\Delta$ $\ell$ / $\ell$$_0$).
The calculated $\tau$ of x- and y-axis, shown in Table ~\ref{Tab3}, are 0.240 and 0.453~ps
for electrons, while $\tau$ is 0.964 and 0.302~ps for holes, indicating a large
anisotropic transport behavior.

According to the Eq.~(\ref{mobility}), $\sigma$ is inversely probational to the effective mass,
contrary to the situation of previous \textit{S}$_{2D}$. The calculated $\sigma$ of penta-silicene
is shown in Figure~\ref{fig2}c,d. Overall, $\sigma$ significantly increases as a function of
\textit{n}. $\sigma$ of 300 K and 600 K are comparable when \textit{n} $<$ 10$^{13}$ cm$^{-2}$.
However, $\sigma$ of 300 K increases faster than $\sigma$ of 600 K when \textit{n} $>$ 10$^{13}$ cm$^{-2}$.

According to Eq.~(\ref{electronic_kaapa2}), the electronic thermal conductivity $\kappa_e$
can be obtained based on $\sigma$, shown in Figure~\ref{fig3}a,b. In semiconductors, phonons,
especially acoustic phonons, are the main carrier of heat transport. Hence, at low
concentration (\textit{n} $<$ 10$^{12}$ cm$^{-2}$), $\kappa_e$ has a value of around 0.2~W/mK.
As \textit{n} increases, penta-silicene is increasingly doped due to the augment of corresponding
$\sigma$. For a larger \textit{n}, $\kappa_e$ can not be neglected since more holes (electrons)
are doped and the penta-silicene is more like a good conductor\cite{madsen2006,gao2019ultralow}.

Generally, \textit{S} decreases and $\sigma$ increases as a function of carrier concentration.
An admirable thermoelectric material needs large \textit{S} and $\sigma$ simultaneously. Power
factor \textit{PF} (\textit{S$^2$}$\sigma$) is a good indicator to describe this joint effect
of \textit{S} and $\sigma$. The calculated \textit{PF} is shown in Figure~\ref{fig3}c,d.
It increases at low concentration due to the enhancement of \textit{S} and
$\sigma$. Then \textit{PF} decreases after climbing to the top since \textit{S} is suppressed
at elevated doping. For p-type, \textit{PF} along the x-axis is larger than that of the y-axis.
However, n-type is the opposite situation (\textit{PF}$_y$ $>$ \textit{PF}$_x$). At 300 K, the
maximum \textit{PF} for p-type doping is 156.40~mW/mK$^2$ and 40.23~mW/mK$^2$ at
5.09 $\times$ 10$^{13}$ cm$^{-2}$ concentration along the x- and y-axis, respectively.
Similarly, maximum \textit{PF} for n-type doping is 61.75~mW/mK$^2$ and 39.22~mW/mK$^2$ at
1.55 $\times$ 10$^{13}$ cm$^{-2}$ concentration along both axis at the same temperature. Such
\textit{PF} values are quite larger than 2D tellurium (57.3~mW/mK$^2$)\cite{gao2018high} and
2D SnSe (57.3~mW/mK$^2$)\cite{chang20183d}.

A good thermoelectric material should satisfy that the distribution of electronic energy carriers
at Fermi level is as narrow as possible, as well as high carrier velocity according to the Mahan's
guideline\cite{mahan1996best}. Subsequently, researchers find an effective approach to enhance
Seebeck \textit{S} and electrical conductivity $\sigma$ simultaneously, where the electronic bands
around \textit{E}$_F$ contain both flat and dispersive bands in the dependent momentum space. It is
known as a ``pudding-mold'' band structure\cite{kuroki2007pudding,usui2013large}. Flat bands increase
the DOS and dispersive bands induce high carrier velocities. It has been used to explain many
prominent thermoelectric performances, such as Na$_x$CoO$_2$\cite{kuroki2007pudding},
SnSe\cite{zhao2014}, PbTe$_{1-x}$Se$_x$\cite{pei2011convergence}, and tellurium\cite{gao2018high}.
The ultrahigh \textit{PF} of penta-silicene is derived from four
hole pockets and relatively flat bands with the same spirit of ``pudding-mold'' shown in
Figure~\ref{fig2}c,d.

\subsection*{Lattice thermal conductivity}
Maybe the lattice thermal conductivity $\kappa_L$ is the only parameter that can be
tuned independently. The calculated phonon transport property of penta-silicene is
shown in Figure~\ref{fig4}. The phonon dispersion in Figure~\ref{fig4}a is free from
imaginary number, indicating a strong dynamical stability of penta-silicene. The
sound velocity is defined as $\upsilon = \frac{\partial \omega} {\partial q} \mid_{q=0} $.
The calculated $\upsilon$ of acoustic phonon modes TA and LA are shown in
Table ~\ref{Tab3} (ZA mode is a quadratic function of wave vector).
More details about this parabolic ZA mode can be found in the Supporting Information.
$\upsilon_{TA}$ is
3.99~km/s and 4.70~km/s along the x- and y-axis. $\upsilon_{LA}$ is larger than
$\upsilon_{TA}$, having values of 4.77~km/s and 5.54~km/s along with both directions.
$\kappa_L$ as a function of temperature is shown in Figure~\ref{fig4}b. At room
temperature, $\kappa_L$ penta-silicene is 1.66~W/mK and 1.29~W/mK along x- and y-axis.
There are two types of scatterings of phonon transport. One is the Normal process and
the other one is the Umklapp (U) process. The latter is the only contributor to the
final thermal resistance. The touchstone to explore the role of the U process is to
study $\kappa_L$ as a function of temperature. If $ \kappa_L \propto \frac{1} {T} $,
the U process dominates the heat transport in this material. Figure~\ref{fig4}b of
penta-silicene shows a good example of this type in heat transport behavior.

\begin{figure*}[t!]
\includegraphics[width=2.0\columnwidth]{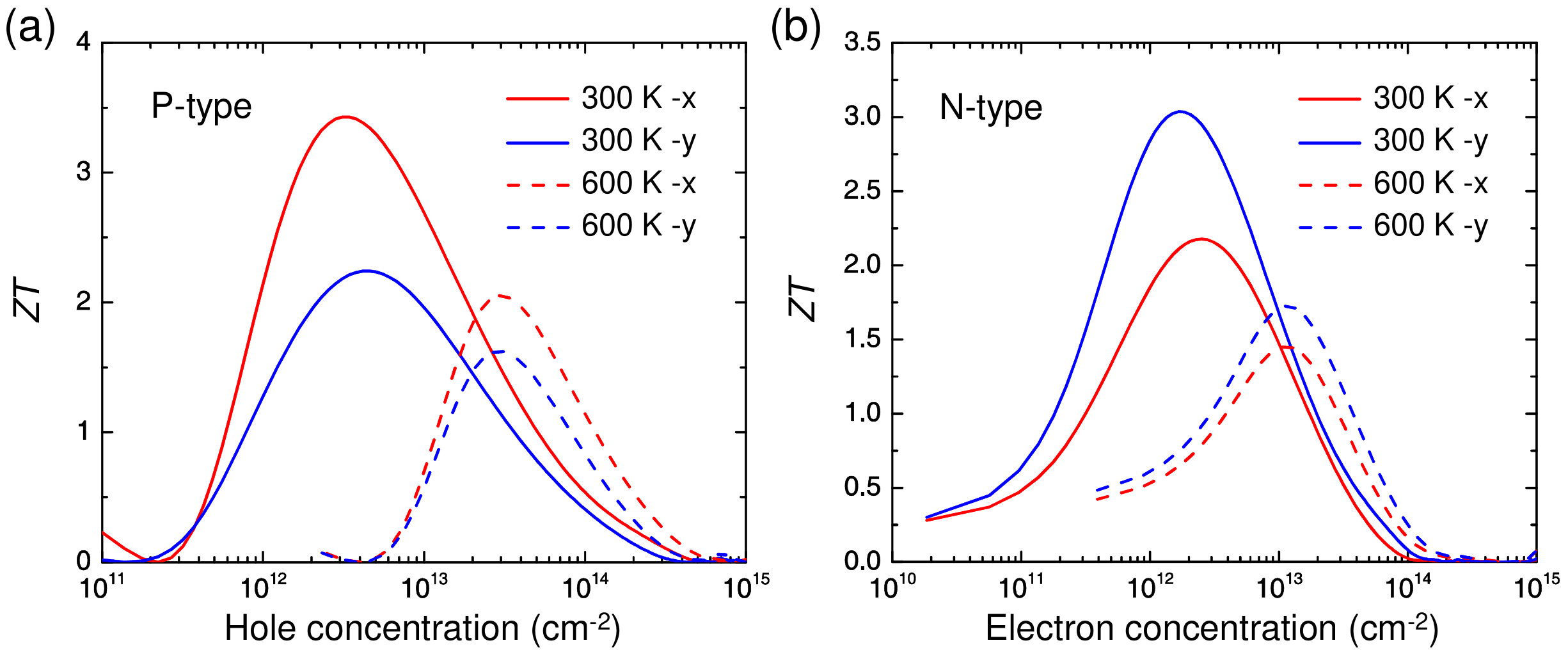}
\caption{Calculated figure of merit \textit{ZT} of (a) p-type and
(b) n-type penta-silicene as a function of the carrier concentration
along x- and y-axis at 300 K and 600 K.}\label{fig5}
\end{figure*}

According to the Slack model, a small $\Theta_D$ generally means a small $\kappa_L$
since acoustic phonons are the main carrier in heat transport. The definition is
$\Theta_D = h \omega_{max}/k_B$ in which $\omega_{max}$ is the maximum
frequency of the acoustic phonon branches. The calculated $\Theta_D$ of penta-silicene
is shown in Table ~\ref{Tab3}. The largest $\Theta_{LA}$ has a value of 131.51~K
along the x-axis. This value is about one-fifth of
penta-graphene ($\Theta_{LA}$ = 692~K)\cite{wang2016lattice}. What's more, the
frequency-resolved $\kappa_L$ of penta-silicene further confirms this discussion.
Phonon vibrations below 50~cm$^{-1}$ contribute most to the final $\kappa_L$
and heat transport whether it is the x-axis or the y-axis. Phonon relaxation time
in Figure~\ref{fig4}d also verifies this fact. In the three branches of acoustic
phonons, The LA branch has the largest phonon relaxation time compared with the
ZA and TA modes. The ultralow $\kappa_L$ of penta-silicene originates from the
weak phonon harmonic interaction and strong anharmonic
scattering\cite{gao2019ultralow,bao2018review,ouyang2018thermal}. We hope this low
$\kappa_L$ would induce a high figure of merit \textit{ZT} of penta-silicene.

\subsection*{Figure of merit \textit{ZT}}
Based on the above results of electronic and phonon transport properties, one can
evaluate the figure of merit \textit{ZT} written as
\begin{equation} %
\label{ZT}
ZT = \frac{ S^2 \sigma } {\kappa_e + \kappa_L } T. \\
\end{equation}

The calculated \textit{ZT} of penta-silicene for both p-type and n-type is shown
in Figure~\ref{fig5}. Interestingly, due to the high \textit{PF} and low $\kappa_L$,
penta-silicene shows a large \textit{ZT} ($>$ 2) at room temperature for both x-
and y-axis. Specifically, maximum \textit{ZT}$_h$ (for the hole) has values of 3.43
and 2.24 at 3.42 $\times$ 10$^{12}$ cm$^{-2}$ and 4.60 $\times$ 10$^{12}$ cm$^{-2}$
concentrations along x- and y-axis at room temperature. Similarly, maximum
\textit{ZT}$_e$ (for the electron) has values of 2.18 and 3.04 at
2.51 $\times$ 10$^{12}$ cm$^{-2}$ and 1.85 $\times$ 10$^{12}$ cm$^{-2}$ along
x-and y-axis at 300 K. In a graphene-based field-effect transistor,
a doping level up to 4 $\times$ 10$^{14}$ cm$^{-2}$ for both electrons and holes
has been reached by electrical gating\cite{efetov2010controlling} and ionic liquid
injection\cite{chuang2014high}. Due to the portability of the experimental
technique, 2D material, like graphene, MoS$_2$, and black phosphorus, generally
can reach 4 $\times$ 10$^{14}$ cm$^{-2}$ for both electrons and holes doping level.
Note that our carrier concentrations ($<$ 4 $\times$ 10$^{14}$ cm$^{-2}$) all are
reachable in these current experimental technique.
%
Therefore, our result indicates that monolayer penta-silicene is a promising
thermoelectric material.

\section*{Conclusion and Discussion}
According to Eq.~(\ref{ZT}), the figure of merit \textit{ZT} is closely related to
the temperature. Generally, \textit{ZT} increases when temperature increases, such as
SnSe with \textit{ZT} of 2.6 $\pm$ 0.3 at 923~K\cite{zhao2014}
and PbTe–PbS pseudo-binary with
\textit{ZT} of 2.3 at 923~K\cite{wu2015superior}. Taking into account the application
of daily life, near room temperature thermoelectric materials with high \textit{ZT}
are desirable. There is no doubt that Bi$_2$Te$_3$-based thermoelectric materials
still dominate around room temperature. However, the tellurium element has very low
reserves on earth and much expensive. How to find inexpensive thermoelectric material
that works efficiently at room temperature is still an open question. Very recently,
some researchers have made progress in this direction. For example,
Mg$_3$Bi$_2$ alloy has a \textit{ZT} of
0.9 at 350~K\cite{mao2019high}. Penta-silicene with \textit{ZT} of 3.43 probably
enriches the room temperature thermoelectric materials. Besides, how to manipulate
defects, lattice symmetry, spin, and electron-phonon coupling to further enhance
room temperature \textit{ZT} is one of the cutting edges in the thermoelectric field.

For 2D materials, stability is the most important before the property. As a matter
of fact, metastable phases are quite common in condensed matter. Materials science
overwhelmingly deals with metastable states. Besides penta-silicene, many simple
light-element compounds including most hydrocarbons, nitrogen oxides, hydrides,
carbides, carbon monoxide (CO), alcohols and glycerin—are also metastable at
ambient condition. Nevertheless, the ubiquitous metastable phases do not produce
a bad effect on their vast and tremendous applications in modern industrial societies
and our daily life. The world has been varied and diverse due to the fact that the
metastable states provide the complexity of structures and energy transformation.
Furthermore, previous works have successfully obtained penta-silicene nanoribbon on
Ag(110) with exotic electronic
properties\cite{cerda2016unveiling,prevot2016si,sheng2018pentagonal,guo2019lattice}.
At the energy level, the cohesive energy \textit{E}$_c$ of penta-silicene is 3.92~eV/atom
that is comparable with the 3.71~eV/atom\cite{fleurence2012experimental,feng2012evidence}
of silicene (hexagonal symmetry with 2 atoms in the primitive cell) and
3.61~eV/atom\cite{liu2014phosphorene,li2014black} of black phosphorene. Note that above
two metastable 2D materials, at present, can be easily obtained in experiment based
on the advanced experimental technique such as creating a complex chemical environment
and variable substrate effect. Besides, we also calculate other two allotropes
of penta-silicene. There are Si-\textit{Cmma} and Si-\textit{Pmma}\cite{zhou2019si}. The
calculated \textit{E}$_c$ of Si-\textit{Cmma} and Si-\textit{Pmma} are 4.17 eV/atom and
4.18 eV/atom, relatively more stable than penta-silicene.

Moreover, we offer three very promising experimental approaches to make free
standing penta-silicene potentially. Besides, these methods also have great potential for
other 2D materials. ``Geometrical frustration'' was first raised by Joel Therrien who are
synthesizing the penta-graphene and other free-standing carbon rings, such as the
U-carbon\cite{gibbs2019new}. As far as we can see, penta-silicene has no counterpart
in the corresponding bulk material. This fact is quite different from graphene and
graphite. Crystalline oxide perovskites, like penta-silicene, also cannot be obtained
by mechanical exfoliation. However, monolayer freestanding crystalline oxide perovskites
has been grown by ``reactive molecular beam epitaxy''\cite{ji2019freestanding}. Besides,
free-standing monolayer amorphous carbon has successfully been created by ``laser-assisted
chemical vapor deposition'' method\cite{toh2020synthesis}. This monolayer amorphous carbon
contains 5-, 6-, 7-, and 8-membered carbon rings. Note that penta-silicene only consists
of 5-membered silicon rings. Encouraged by these above advanced technologies in the
experimental cutting edge, we conclude monolayer free-standing penta-silicene and other
novel nanostructures can also be obtained on the near horizon.

Sometimes, the figure of merit \textit{ZT} based on the mobility $\mu$  in Eq.~(\ref{mobility})
is a little overestimated since  acoustic phonon is not the only factor to impact the $\mu$
that determines the final \textit{ZT}. As we mentioned before, nonpolar optical phonons, and
ionized impurities also influence $\mu$. Generally, the impact of these two factors can be
neglected and the role of longitudinal acoustic phonons is
dominant\cite{qiao2014,bardeen1950deformation,pei2011convergence,wang2012weak}. Given
that realistic calculation of couplings between nonpolar optical phonons and electrons,
ionized impurities and electrons are beyond our current computing capabilities, they are
interesting open questions that deserve further exploration.

DP theory is based on the rigid band approximation. Surprisingly, this
approximation works well for most cases when the electronic bands around the fermi
level are not highly degenerate\cite{gao2019degenerately,gao2018high}. Besides,
VASP, Quantum ESPRESSO, and BoltzTraP
softwares\cite{kresse1996efficient,kresse1996efficiency,kresse1994ab,madsen2006},
to name a few, also can add electrons into or remove electrons from the system
through a compensating uniform charge background of opposite sign to maintain
charge neutrality, which verifies the correctness of this approximation
additionally.

DP theory has been widely used and has been successful in calculating the
intrinsic (free-standing and defects-free) mobility, which can be found everywhere.
Certainly, almost any theoretical model has assumptions, either large or small,
such as famous density functional theory, Dulong-Petit law (high temperature) and even
Newton's laws of motion (inertial system). DP theory is also included. As a matter of fact,
in the DP theory, only the longitudinal-acoustic phonon is considered without any dispersion
and the electron-phonon matrix is expressed by the DP constant and the elastic constant that
are described in the Eq.~(\ref{mobility}). The optical phonon scatterings are absent in the
DP theory. However, DP theory’s results are qualitatively and even quantitatively reasonable
in many cases compared with the full evaluation of electron-phonon
coupling\cite{nakamura2017intrinsic}. Since this calculation is far beyond our
current computational capabilities, hence, we suggest leaving the e-ph coupling as an open
question, which should be explored further.

How a 2D functional material, like our penta-silicene, is to be used in bulk
devices? There are 4 options to integrate bulk materials with 2D materials for physical
coupling and applications\cite{bae2019integration}. On the one hand, one can construct
2D materials on 3D materials (2D-on-3D) heterostructures. The technique methods are Van
der Waals epitaxy (vdWE), wet transfer and metal-induced quasi-dry transfer process. On
the other hand, the realization of 3D-to-2D heterostructures is still at a premature stage.
But the relevant experiment is growing fast, which has significantly broadened the material
beyond conventional 3D materials-based heterostructures.

In summary, we have calculated the thermoelectric performance of monolayer penta-silicene
by first principles. It has good thermal, dynamical, and mechanical stability compared with
many other typical 2D materials, such as hexagonal silicene and black phosphorous. Lattice
thermal conductivities of penta-silicene have values of 1.66 and 1.29~W/mK along x- and
y-axis, respectively. Superior electronic properties originate from the ``pudding-mold-like''
shape of valence bands around the Fermi level. Ultralow thermal conductivity and high
power factor collaboratively lead to an ultrahigh \textit{ZT} of penta-silicene. At room
temperature, maximum \textit{ZT} has values of 3.43 and 3.04 for hole doping and electron
doping, respectively. Our work has indicated that penta-silicene is a promising
thermoelectric material, especially near room temperature. Layer--dependent
thermoelectric property of penta-silicene deserves more follow up study in the future.


\quad\\
{\noindent\bf Author Information}\\

{\noindent\bf Corresponding Author}\\
$^*$E-mail: {\tt zhibin.gao@nus.edu.sg} \\

{\noindent\bf ORCID}\\
Zhibin Gao: 0000-0002-6843-381X \\

{\noindent\bf Supporting Information}\\
The Supporting Information is available free of charge on the
ACS Publications website via the Internet at
https://pubs.acs.org/journal/aamick.

Room temperature lattice thermal conductivity of penta-silicene
as a function of the cutoff for the interaction range of
anharmonic force constants and phonon dispersion superposed
on the spectra from continuum theory.\\

{\noindent\bf Notes}\\
The authors declare no competing financial interest.


\begin{acknowledgement}\\
We are grateful to Wu Li for the kind guidance on calculation of 
lattice thermal conductivity. We acknowledge Jinyang Xi for many 
fruitful discussions and good suggestions. We also thank Wen Shi and 
Tianqi Deng for valuable discussions and kind help. We acknowledge 
the financial support from MOE tier 1 funding of NUS Faculty of Science,
Singapore (Grant No. R-144-000-402-114).
\end{acknowledgement}


\providecommand{\latin}[1]{#1}
\makeatletter
\providecommand{\doi}
  {\begingroup\let\do\@makeother\dospecials
  \catcode`\{=1 \catcode`\}=2 \doi@aux}
\providecommand{\doi@aux}[1]{\endgroup\texttt{#1}}
\makeatother
\providecommand*\mcitethebibliography{\thebibliography}
\csname @ifundefined\endcsname{endmcitethebibliography}
  {\let\endmcitethebibliography\endthebibliography}{}

\newpage

\begin{figure*}[tbp]
\begin{center}
\includegraphics[width=6.0in]{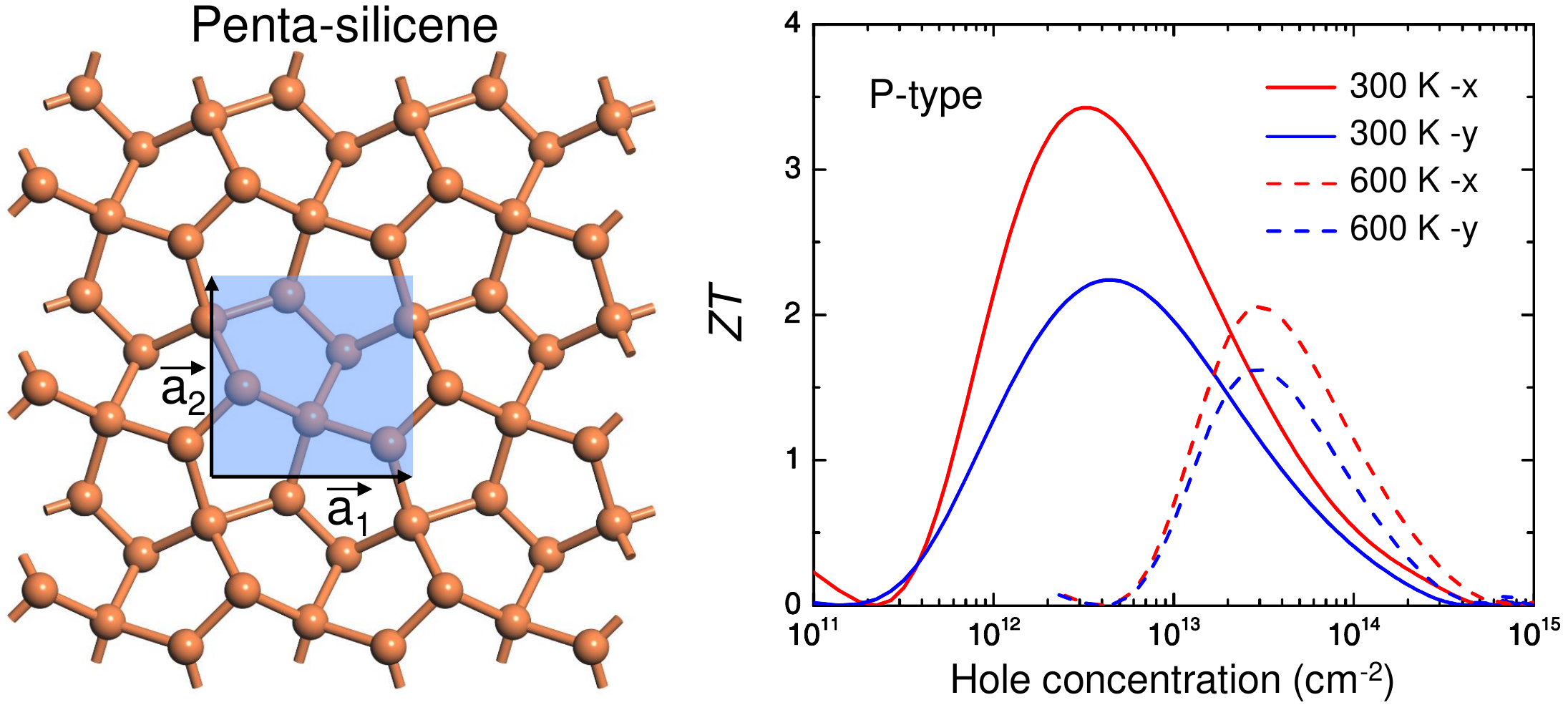}
\captionsetup{labelformat=empty}
\caption{Table of Contents Graphic}
\end{center}
\end{figure*}

\end{document}